\documentclass{aa} % for the letters  
\usepackage{natbib}
\bibpunct{(}{)}{;}{a}{}{,} % to follow the A&A style
\usepackage{graphicx}
\usepackage{caption}
\usepackage{subcaption}
\usepackage{amssymb}
\usepackage{comment}
\usepackage{txfonts}
\usepackage[dvipsnames]{xcolor}
\usepackage[colorlinks=true,citecolor=RoyalBlue, linkcolor=Emerald]{hyperref}

\usepackage[normalem]{ulem}

\begin{document}

   \title{
   A break in planet occurrence near the pebble isolation mass 
   should be observable by the \textit{Roman} microlensing survey 
   }

   \author{Claudia Danti\inst{1}
          \and
          Michiel Lambrechts\inst{1}
          \and 
          Hannah Diamond-Lowe \inst{2}
          }

   \institute{Center for Star and Planet Formation, Globe Institute, Øster Voldgade 5, 1350 Copenhagen, Denmark
   \and
   Space Telescope Science Institute, 3700 San Martin Drive, Baltimore, MD 21218, USA \\
   \email{claudia.danti@sund.ku.dk, danticlaudia@gmail.com}
    }

   \date{}

 \abstract
   {
% state-of-the-art (context)
Microlensing detections are uniquely well-suited to probing the population of planets outside the water iceline, down to planetary masses comparable to the Earth.
% methods
Here, we perform 1D pebble-accretion population synthesis simulations to explore a sample of iceline planets around stars with masses and metallicities similar to the target population of the Galactic Bulge Time-domain microlensing survey of the \textit{Nancy Grace Roman Space Telescope}.
%what we find
We find that the planet distribution in the microlensing sensitivity space deviates from a log-uniform distribution in mass and orbital radius. 
When planetary core growth comes to a halt as planets reach the pebble isolation mass, $M_{\mathrm{iso}}$, the combined effects of planetary migration and runaway gas accretion create an occurrence break. Our simulations highlight that, between $1$ and $50$ AU, the fraction of stars hosting isolation-mass planets ($1$ to $5$ $M_{\mathrm{iso}}$)
is lower by a factor $20$ compared to less massive planets ($0.2$ to $1$ $M_{\mathrm{iso}}$). 
If this break in planetary occurrence rates around the pebble isolation mass is detected in future lensing surveys, it would further validate the core accretion paradigm for giant planet formation.
}

\authorrunning{C. Danti et al.}
\titlerunning{A break in planet occurrence near the pebble isolation mass
should be observable by microlensing surveys}
   \maketitle
%
%-------------------------------------------------------------------

\section{Introduction}
Microlensing surveys are unique in their ability to detect sub-Earth mass planets beyond the water iceline \citep{zhu_2021}.
The upcoming Galactic Bulge Time-Domain Survey by the Nancy Grace Roman Space Telescope (\textit{Roman}) will be sensitive to $0.02$\,$M_{\oplus}$ planets at orbital distances of $4$\,AU, using the microlensing detection method, and is expected to yield around a thousand new lensing exoplanets \citep{Penny_2019}.
%  description of microlensing
The microlensing detection technique is based on the detection of a magnification event \citep{gaudi_review_2012}. This increase, or decrease, of the light curve of a background source star, typically at a distance of $\approx$\,$8$\,kpc in the galactic bulge, is due to the gravitational perturbation of  a foreground lens star, typically at a distance of $\approx$\,$4$ kpc.
The presence of a planet around the lens star then creates small planetary caustics that induce variations on the single lens magnification event \citep{paczynski_1986}.
The typical host stars of lensing planets are M-dwarfs, due to their 
higher occurrence in the galactic disc compared to solar-like stars \citep{Chabrier_2005}.

% microlensing surveys and Roman
Previous work has argued that planet formation is efficient around M-dwarf stars inside of the water iceline in the pebble accretion framework \citep{Liu_2020,mulders_why_2021,chachan_small_2023}.
Indeed, the exoplanet population synthesis model of \citet{Pan_2025} finds that super-Earths occurrence rate peaks around early M-dwarfs ($\approx 0.5 \: M_{\odot}$) and decreases towards late M-dwarfs ($0.1- 0.2 \: M_{\odot}$) and solar-type stars ($0.6-1 \: M_{\odot}$).
In contrast, warm and cold giants show a monotonically increasing occurrence rate with stellar mass. 
These theoretical findings appear to be consistent with observed occurrence rate trends with stellar masses \citep{mulders_2015_b, Hsu_2020, Sabotta_2021}.
Building on these works, we expand our previous exoplanet synthesis model \citep{Danti_2025} to investigate planet formation in the sensitivity domain of lensing surveys, around the iceline of a galactic stellar population.

\section{Model}
\label{sect:model}

%  PLANET FORM MODEL 
To perform the population synthesis simulations, we used a 1D pebble accretion model \citep{Danti_2025}
, with adjustment to account for stellar mass dependencies (Appendix\,\ref{app:stellar_mass_dependencies}).
%    PLANETESIMALS
The planetary embryo's seeds, taken from the top of the streaming instability distribution \citep{Liu_2020}, grow by accreting pebbles \citep{ormel_effect_2010, lambrechts_rapid_2012} that are both drift- and fragmentation-limited in size \citep{brauer_coagulation_2008}. 
% MIGRATE
The planetary embryos migrate through the disc in the type I migration regime \citep{paardekooper_torque_2010}. Once the protoplanets reach isolation mass, they stop accreting solids \citep[][see also Appendix\,\ref{app:pebiso}]{lambrechts_separating_2014}. Subsequently, the planets switch to type II migration \citep{kanagawa_radial_2018} and start accreting gas.
% GAS ACC
The gas accretion phase is characterised by an initial slow Kevin-Helmholtz contraction \citep{Ikoma_formation_2000}, followed by runaway gas accretion \citep{Tanigawa_final_2016}, limited by the accretion rate through the gap that equals the gas accretion rate onto the central star \citep{lubow_accretion_2006}.
The total amount of material available for solid accretion is given by a flux of pebbles set in the outer disc as a fraction $Z$, corresponding to the stellar dust-to-gas ratio, of the gas accretion rate onto the central star \citep{johansen_how_2019}. 
%and,
In case of multiple embryos being present (see Appendix \ref{app:multiple_planets}), the pebble flux is reduced by mutual pebble filtering \citep{guillot_filtering_2014}.

%  DISC 
We use a disc model with a prescribed gas accretion rate, following \citet{Liu_2020}, and a temperature profile that can be entirely irradiated \citep{ida_radial_2016} or include moderate accretion heating \citep[$\epsilon_{\rm el}=10^{-2}$, see Appendix A in][]{Danti_2025}.
The disc viscosity is fixed at $\alpha_{\nu} = 10^{-2}$ \citep{appelgren_disc_2023}, while the turbulent stirring and fragmentation parameters are $\alpha_{z} = \alpha_{\rm frag} = 10^{-4}$, representing a low-turbulence midplane as supported by recent observations of dust scale heights \citep{pinte_2016} and fragmentation-limited particle sizes \citep{jiang_grain-size_2024}.
The inner edge of the disc is set to be the inner magnetospheric cavity 
\citep{liu_dynamical_2017}.

\section{Results}
\label{sect:results}
\subsection{
Dependency of the synthetic exoplanet population on disc mass budget for typical lensing stars
}
\label{sect:pop_synth_stars}
An important uncertainty in our model is the poorly known scaling relation between stellar mass and the gas accretion rate onto the star, 
\begin{align}
    \label{eq:mdot_linear}
    \dot{M}_{\star}(t,M_{\star}) = \dot{M}_{\rm H16}(t) \left(\frac{M_{\star}}{0.7M_{\odot}}\right)^{\beta}.
\end{align}
Here $\dot{M}_{\rm H16}$ is the \citet{hartmann_accretion_2016} gas accretion rate observational fit anchored at $M_{\star}$=$0.7\,M_{\odot}$(their Eq.\,12).
The linear scaling ($\beta$=$1$) and quadratic scaling ($\beta$=$2$), represent, respectively, the weakest and steepest $M_{\star}$ dependencies
that encompass the observational spread \citep{Alcala_2014, manara_2012, Venuti_2014}.
We also explore a steeper-than-linear relation based on the observation of a large sample of sources in the Orion Nebula Cluster 
\citep[][]{manara_2012} in Appendix \ref{app:stellar_acc_rate}.
We therefore simulate a set of $2000$ single planets around a typical lensing star, $M_{\star} \approx 0.2 \: M_{\odot}$ \citep{Bochanski_2010, Henry_2024}, to explore the effects of the different gas accretion rate scalings with stellar mass, assuming either the $\beta$=$1$ or $\beta$=$2$ limit case (Eq.\,\ref{eq:mdot_linear}).

\begin{figure}[t!]
    \centering
    \includegraphics[width=\linewidth]{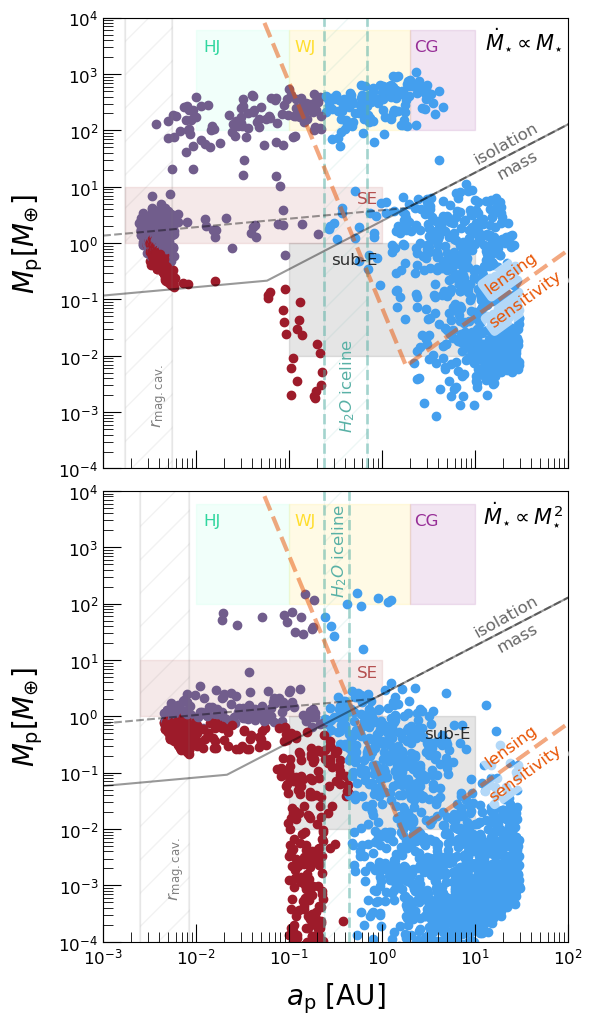}
    \caption{
    Mass and semi-major axis of the synthetic populations of planets around $0.2 \: M_{\odot}$-stars, with our fiducial mass budget (Eq.\,\ref{eq:mdot_linear} with $\beta$=$1$, top panel) and a lower mass budget 
    (Eq.\,\ref{eq:mdot_linear} with $\beta$=$2$, bottom panel).
    The colours represent icy embryos with initial and final position outside the iceline (blue), mixed embryos that start outside and migrate inside the iceline (purple) and rocky embryos with initial and final position inside the iceline (red). 
    The pebble isolation mass at initial and final simulation time is marked by the grey dashed and solid line, respectively.
    The lensing sensitivity region is marked with the orange dashed lines, assuming the lens star is in the disc ($4$\,kpc) and the source star in the bulge ($8$\,kpc). 
    A quadratic scaling of the accretion rate with stellar mass (bottom panel) hinders the formation of planets outside the iceline as well as gas giants.
    }
    \label{fig:pop_synth}
\end{figure}

For these synthetic populations, the stellar metallicities are sampled from Gaussian distributions peaked in $\mu_{\rm [Fe/H]} = -0.02$ with spread $\sigma_{\rm [Fe/H]} = 0.22$,  following spectroscopic surveys of solar neighbour stars \citep{santos_2005, Emsenhuber_2021}. 
The initial dust-to-gas ratio of the disc, $Z$, is then tied to the stellar metallicity through $Z = 10^{\rm [Fe/H]} f_{\rm dtg, \odot}$, with $f_{\rm dtg, \odot} = 0.0149$ solar dust-to-gas ratio \citep{Lodders_2003}, where we assumed that the stellar metallicity distribution is independent of the stellar mass \citep{santos_2003}.
The initial position of the embryos are randomly drawn from a log-uniform distribution between $0.1$ and $30$ AU.
Finally, the initial disc lifetime is randomly sampled from a normal distribution with a mean lifetime of $\mu_{\tau } = 5 \: \rm Myr$ and with a spread of $\sigma_{\tau}=0.5 \: \rm Myr$. 
The resulting synthetic populations are shown in Figure \ref{fig:pop_synth}, for a linear gas accretion scaling (top panel) and a quadratic one (bottom panel). 
The coloured boxes define the different types of planets, whose definitions can be found in Table \ref{tab:planets_def}. The lensing sensitivity region, parametrised through Eq. \ref{eq:s_q}, is marked by the orange dashed line.

The linear scaling of gas accretion rate yields a population of $\approx 24 \%$ super-Earths and $\approx 10 \%$ gas giants, roughly consistent with the lower limit of super-Earth occurrence $\eta_{\rm SE} \approx 55_{-26}^{+40} \%$ and the upper limit of giant planet occurrence $\eta_{\rm GP} \approx6_{-3}^{+4} \%$ around M-dwarfs \citep{Sabotta_2021}.
In contrast, the quadratic scaling with accretion rate significantly hinders the growth of embryos, especially outside the iceline, resulting in discrepant occurrence fractions with respect to the observed exoplanet census.

For completeness, we also show in Appendix\,\ref{app:stellar_acc_rate} that the observationally-motivated steeper-than-linear relation by \citet{manara_2012} similarly leads to small disc mass budgets in our model that under-predict the occurrence rate of super-Earths and gas giants. Given this, and the known large observational spread and uncertainties on stellar ages \citep{hartmann_accretion_2016}, we choose for simplicity a linear $\beta$=$1$ scaling.

\begin{figure*}[t!]
    \centering
    \includegraphics[width=\textwidth]{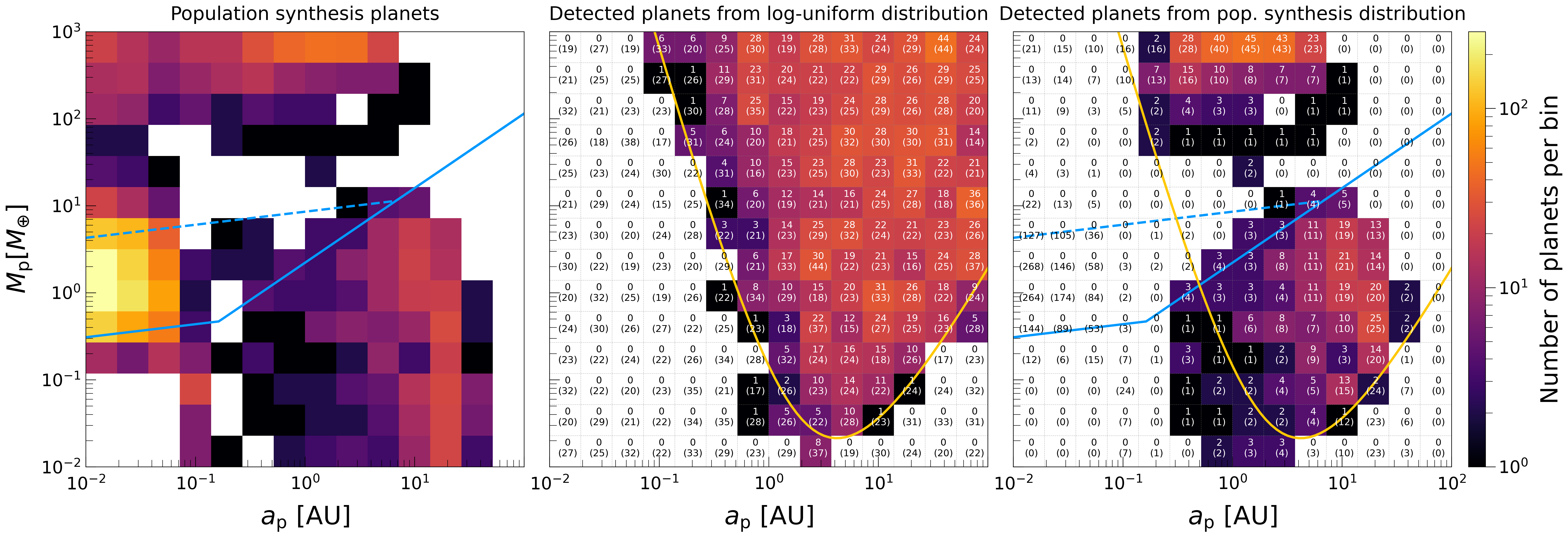}
    \caption{
    \textbf{Left panel}: Synthetic exoplanet population, colour coded by number of planets per bin in logarithmic semi-major axis and planet mass space, using the nominal model for a sample of $N_{\rm star}$=$5000$ different mass stars (Section \ref{sec:planyield}). 
    For reference, we show the pebble isolation mass as a function of orbital radius, for a $0.2 \: M_{\odot}$ star, at the initial (blue dashed) and final simulation time (solid blue lines).
    \textbf{Central panel}: Total number of detected lensing planets per bin, assuming an underlying log-uniform distribution 
    of $5000$ planets 
    in semi-major axis and mass, with the 
    lensing detection efficiency specified in Section \ref{sec:planyield}. 
    Annotated in each bin is the number of planets detected per bin, with the underlying number of planets in that bin in parentheses.
    The solid orange line shows the $3$-planet detection line for the \textit{Roman} Space Telescope \citep{Penny_2019}.
    \textbf{Right panel}: Total number of detected lensing planets per bin, using the synthetic population shown in the left panel. 
    The dearth of planets with masses near the pebble isolation mass at the approximate location of the water iceline is located in the high-sensitivity region of lensing surveys.
    }
    \label{fig:lensing}
\end{figure*}

\subsection{Planetary yields for a stellar lensing population}
\label{sec:planyield}

We now present our synthetic planet population drawn from $N_{\star}=5000$ stars sampling the IMF distribution of \citet{Chabrier_2005}, with stellar masses ranging from $0.01$ to $5 \: M_{\odot}$, assuming each star hosts a single planet.
We calibrate the key parameters of the synthesis model such that we reproduce the observed super-Earth (SE)
occurrence around FGK stars, $P(\rm SE) \approx 30-50 \%$ \citep{zhu_about_2018}. 
The nominal model  yields an occurrence of $P(\rm SE) \approx 35 \%$ when assuming
(i) a linear scaling of gas accretion rate (Eq. \ref{eq:mdot_linear}, see Figure \ref{fig:pop_synth}, top panel),
(ii) a particle fragmentation velocity of $10 \: \rm m/s$ outside the iceline and $1 \: \rm m/s$ inside the iceline,
(iii) a moderate accretion-heated disc (see Appendix \ref{app:nominal_model_calibration} for the model calibration details and description).
The left panel of Figure \ref{fig:lensing} shows a break in occurrence above the pebble isolation mass (blue curves).
This is a direct consequence of the combined effects of planet migration and runaway gas accretion, because planets that do make it beyond isolation mass either migrate all the way to the inner disc or tend to grow into more massive gas giants.
Therefore, the region above the pebble isolation around the iceline ($\approx$ $1$ AU) is depleted in planets.

To provide a planetary yield from our synthetic population of planets, we first need to estimate the detection sensitivity of a microlensing survey. 
The perturbation of the observed light curve due to the presence of a planet is strongest if the planet lies near the Einstein radius 
(at projected separation $s=a_{\rm p}/R_{\rm E}=1$, where $a_{\rm p}$ is the planet's semi-major axis and $R_{\rm E}$ is the Einstein radius), with a decreasing efficiency at smaller ($s<1$) and larger ($s>1$) separations, and an increasing  efficiency for larger planet-to-host star mass ratios ($q = M_{\rm p}/M_{\star}$).
In the case of a high-magnification event, this almost always results in a triangle-shaped detection efficiency in  $(q, s)$ space that, following \citet{gould_2010}, can be parametrised as 
\begin{equation}
\label{eq:s_q}
q =
\begin{cases}
   q_{\mathrm{min}} \, s^{-1/\eta_{-}} & \text{when } s<1, \\
   q_{\mathrm{min}} \, s^{1/\eta_{+}}  & \text{when } s>1,
\end{cases}
\end{equation}
where $\eta_{\pm}$ are the left and right slopes of the triangle
and $q_{\rm min}$ is the minimal detectable planetary mass ratio, corresponding to the bottom tip of the triangle, as illustrated in orange in Figure \ref{fig:pop_synth}.
In our model we assume $100 \%$ detection efficiency inside this triangle and $0 \%$ outside, roughly consistent with the detection efficiency triangles of high magnification events \citep[][]{gould_2010}, but slightly overestimating the detection efficiency in the tail of low-magnification events \citep[][see also Appendix \ref{app:triangle_efficiency} for the lensing sensitivity triangle description]{gould_2010,suzuki_microlensing_2016}.

To model our lensing detection efficiency, we choose parameters broadly consistent with the \textit{Roman} detection efficiency characterised in \citet{Penny_2019}.  
We parametrised the sensitivity triangle in the $(q,s)$ space through Eq. \ref{eq:s_q}, with $q_{\rm min} = 10^{-7}$, and two different slopes for the left and right side $\eta_{-} = 0.25$ and $\eta_{+} = 0.85$ (see left panel of Figure \ref{fig:sensitivity_star_IMF} in Appendix \ref{app:triangle_efficiency}). 
High-magnification events have a sensitivity which is nearly symmetrical in $\log s$, implying $\eta_{-}= \eta_{+}$ \citep{gould_2010, choi_2012b}, while more modest magnification events show higher sensitivities for planets with $s>1$ and lower for planets with $s<1$ \citep{suzuki_microlensing_2016}. This is because planets outside the Einstein ring ($s>1$) cause perturbations in the higher-magnification major image created by the lens star, while planets inside the Einstein ring perturb the already lower-magnification minor image.

The central panel of Figure \ref{fig:lensing} shows the total number of detected planets, assuming one planet per star and an underlying log-uniform distribution of planets in semi-major axis and mass. This is obtained by multiplying the number of planets per mass-position bin by the detection efficiency in that specific bin.
This figure shows that our simplified parametrization qualitatively reproduces the reported detection efficiency of the Roman Space Telescope, assuming the same log-uniform planet distribution \citep[Figure 9 of][]{Penny_2019}. 
This is also reflected in the decline of the detection efficiency (dark violet bins in the central panel of Fig. \ref{fig:lensing}) that follows the so-called three-planet detection line predicted for \textit{Roman} \citep[light blue curve in Fig. 9 in][]{Penny_2019}.

A planet distribution that is log-uniform in position and mass is not supported by our synthetic population (left panel of Figure \ref{fig:lensing}). Moreover, several observational studies have already shown, for example, a decreasing occurrence rate of gas giants at large orbital distances \citep{fulton_california_2021} and gaps in the occurrence of inner disc planets with approximately 2 Earth radii \citep{Fulton_carlifornia_2017}.
Assuming our synthetic population as the underlying distribution (left panel Figure \ref{fig:lensing}), the right panel of Figure \ref{fig:lensing} shows the expected planet yield in a \textit{Roman}-like microlensing survey.
Firstly, we note the presence of a population of detectable ice worlds, planets with masses between $1$ and $10$\,M$_{\oplus}$ outside the water iceline , that have not yet been observed \citep{bitsch_growth_2015}. 
Importantly, we recover in observational space the
reduction in the number of detected planets above the pebble isolation mass, in the range $[1-5] \: M_{\mathrm{iso}}$, of a factor $20$ with respect to lower-mass planets in the range $[0.2-1] \: M_{\mathrm{iso}}$, for semi-major axes between $1$ and $50$ AU, and an increase in giant planets occurrence.
We therefore argue that a lensing survey, like the Galactic Bulge Time-domain survey of the Roman Space Telescope, 
should be able to detect this occurrence break.
If correct, this would reduce the overall lensing yield of the \textit{Roman} survey near this isolation break, compared to an assumed log-uniform mass and radius occurrence. At the same time, it would support the core accretion paradigm for gas giant formation, as a two-step process consisting of the formation of solid migrating cores, followed by a rapid epoch of gas accretion. If the break is not observed, then further studies on the gas accretion mechanisms that shape giant planet formation need to be conducted.

\section{Future work}
A depletion in the occurrence of planets above a critical core mass is a robust feature of population synthesis models, both in planetesimal-based \citep{Ida_2004, Emsenhuber_2021, Emsenhuber_2025} and pebble-based \citep{Bitsch_2018,liu_super-earth_2019} frameworks, even when using MHD wind-driven, low-viscosity discs \citep{weder_2026}. Observationally, some studies hint at the presence of such a depletion feature \citep{Mayor_2011, Bertaux_2022, zang_2025}, while previous ground-based lensing surveys did not \citep{suzuki_2018, Bennett_2021}. However, in a recent lensing study, \citet{zang_2025} argue for an indication of an occurrence break outside the water iceline. Clearly, a future larger lensing sample, like the one provided by \textit{Roman} will be key to confirm or deny its presence.

Several aspects in this study deserve further work to explore the robustness of our findings. 
Importantly, our results are sensitive to the disc model choice (see Section \ref{sect:pop_synth_stars}), which should be further explored, especially in the context of sub-solar stars \citep{rodriguez_2025}.
We use conventional migration prescriptions, but recent works highlight potential outward migration regimes for planets in the Earth-mass regime \citep{Liu_2015, Nielsen_2025} and more massive gas giants \citep{2sanchez_2025b}. 
Moreover, our nominal model does not directly address planetary multiplicity, important for occurrence rate calculations. Although the role of mutual pebble filtering is minor (Appendix \ref{app:multiple_planets}), N-body studies show that collisional growth of embryos can drive growth beyond the pebble isolation mass \citep{lambrechts_formation_2019,sanchez_2025}.
Ideally, these explorations would be conducted in a full population synthesis context \citep{Burn_2021,Pan_2025,Chen_2025} to further test key model assumptions and report theory-motivated planet yields for future lensing surveys.

\begin{acknowledgements}
We thank the anonymous referee for the helpful comments that improved the quality of this manuscript.
C.D. thanks Lizxandra Flores-Rivera, Vera Sparrman and Adrien Houge for helpful discussion and Allison Youngblood for hosting the visit at NASA Goddard Space Flight Center. M.L. acknowledges the ERC starting grant 101041466 EXODOSS. 
\end{acknowledgements}

\bibliographystyle{aa}
%\bibliography{bibliography}

% --------------------------------------------------------- 
\begin{appendix}

\section{Stellar mass dependencies}
\label{app:stellar_mass_dependencies}
\subsection{Stellar luminosity}
We modelled the stellar mass dependence of the luminosity as a simple power law
\begin{equation}
\label{eq:L_star}
    \left(\frac{L_{\star}}{L_{\odot}}\right) = \left(\frac{M_{\star}}{M_{\odot}}\right)^{3/2},
\end{equation}
as it is shown that young ($< 10$ Myrs) stars have luminosities that scale like $L_{\star} \propto M_{\star}^{1-2}$ before settling on a cubic relationship \citep{Liu_2020}.

\subsection{Stellar accretion rate}
\label{app:stellar_acc_rate}
There have been different attempts to observationally constrain the gas accretion rate as a function of stellar mass. 
A recent spectroscopic surveys, based on observations of different star clusters, report a steeper-than-linear $\dot{M}_{\star}-M_{\star}$ relation, with a 1.6-2.0 slope and 1-2 dex spread \citep[e.g.][]{Alcala_2014, Manara_2017b, Manara_2016, Venuti_2014, Venuti_2019b, hartmann_accretion_2016}. The spread in the power law dependence is likely the result of disc evolutionary processes and the initial condition scaling with stellar mass. 
\citet{manara_2012} provided a fitting formula for the gas accretion rate as a function of time and stellar mass based on observations of sources in the Orion Nebula Cluster
\begin{align}
    \label{eq:mdot_manara}
    \log \left(\frac{\dot{M}_{\star}}{M_{\odot} \rm yr^{-1}}\right) = -5.12-0.46 \log \left(\frac{t}{\rm yr}\right) 
    \notag\\
    -5.75  \log \left(\frac{M_{\star}}{M_{\odot}}\right) +1.17  &\log \left(\frac{t}{\rm yr}\right)  \log \left(\frac{M_{\star}}{M_{\odot}}\right),
\end{align}
with $M_{\star}$ central host mass that ranges between $0.05 \: M_{\odot}$ and $2 \: M_{\odot}$ and $t$ disc age. They point out that given the huge uncertainty on the disc age, this fit is invalid for $t<0.3 \: \rm Myr$.
\begin{figure}
    \centering
    \includegraphics[width=\linewidth]{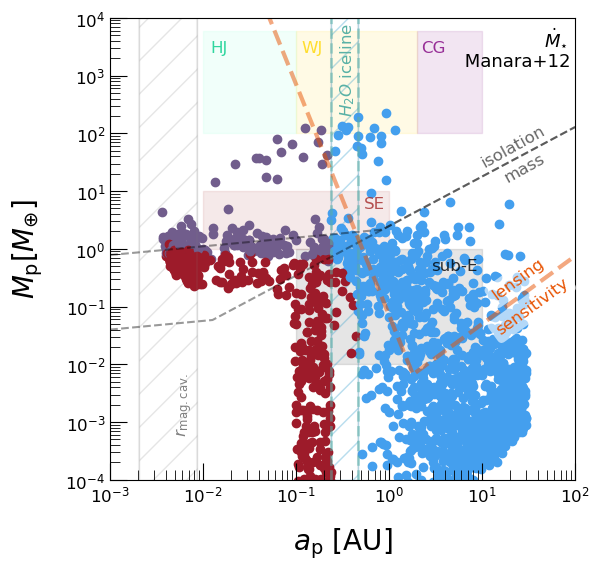}
    \caption{Same as Figure \ref{fig:pop_synth}, but with the steeper-than-linear gas accretion rate of Eq. \ref{eq:mdot_manara} \citep{manara_2012}.}
    \label{fig:liu_gas}
\end{figure}
Figure \ref{fig:liu_gas} shows the same initial seed population around a $0.2 \: M_{\odot}$ star of Figure \ref{fig:pop_synth} using the gas accretion scaling of Eq. \ref{eq:mdot_manara}. This steeper-than-linear relation leads to a lower occurrence of both gas giants and super-Earths, though not as much as the quadratic relation (bottom panel of Figure \ref{fig:pop_synth}).

\subsection{Planetary embryos}
We assume that the planetary embryos have been created through streaming instability within the first $10^{5}$ years of the disc lifetime. We adopt the largest planetesimals formed by streaming instability as the planetary embryo, following \citet{Liu_2020}
\begin{align}
\label{eq:M0_pla}
  M_{0,\mathrm{pla}} = 2\times 10^{-3}
                 \left(\frac{f}{400}\right)
                 \left( \frac{C}{5 \cdot 10^{-5}} \right)
                 \left( \frac{\gamma}{\pi^{-1}} \right)^{a+1}
                    \notag\\    
                 \left(\frac{H/r}{0.05} \right)^{3+b}
                 \left( \frac{M_{\star}}{0.1 \: M_{\odot}} \right)
                 \, M_{\oplus} \, , 
\end{align}
where $f$ is a multiplicative factor that expresses how much the most massive planetesimals exceed the characteristic planetesimal mass of the measured size distribution, and $\gamma$ is the dimensionless gravitational parameter
\begin{equation}
    \label{eq: gamma}
    \gamma = \frac{4 \pi G \rho_{\rm gas}}{\Omega_{\rm K}^2},
\end{equation}
that measures the relative strength between self gravity of the disc and tidal shear. In our simulations, the discs are always non-self-gravitating ($\gamma \neq 1$).
Here we used $f = 400$, $a = 0.5$, $b=0$, and $C=5 \cdot 10^{-5}$ as in \citet{Liu_2020}.

\subsubsection{Pebble isolation mass}
\label{app:pebiso}
Planetary embryos grow in the disc through pebble accretion \citep{ormel_effect_2010, lambrechts_rapid_2012} until they are massive enough to significantly perturb the gas in the disc. The threshold mass for which the gravitational perturbation due to the planet is able to create a pressure bump and halt the accretion of pebbles is called the pebble isolation mass.
It can be expressed as 
\begin{equation}
    \label{eq:peb_iso}
    M_{\mathrm{iso}} \simeq \left(\frac{H}{r}\right)^3 \frac{\partial \ln P}{\partial \ln R} M_{\star} \approx 20 \: \left(\frac{H/r}{0.05}\right)^3 \left(\frac{M_{\star}}{M_{\odot}}\right) M_{\oplus},
\end{equation}
where $H/r$ is the gas disc aspect ratio \citep{johansen_forming_2017}.
In an irradiated disc \citep{ida_radial_2016}, 
the pebble isolation mass is almost independent of the stellar mass
\begin{equation}
    \label{eq:M_iso_irr_star}
    M_{\mathrm{iso, irr}} \approx 2.2 \: \left(\frac{r}{\rm AU}\right)^{6/7} \left(\frac{M_{\star}}{M_{\odot}}\right)^{-1/14} M_{\oplus}\,,
\end{equation}
where we assumed the luminosity $L_\star$ scales with stellar mass as in Eq. \ref{eq:L_star}.
This weak dependency no longer holds in the inner parts of protoplanetary discs, where time-dependent accretion heating can modify the gas scale height \citep[see Appendix A in][]{Danti_2025}.
In this case, the pebble isolation mass is proportional to
\begin{equation}
    \label{eq:M_iso_visc}
    %M_{\mathrm{iso, acc}} \propto
    %\: \left(\frac{r}{\rm AU}\right)^{3/20} \left(\frac{M_{\star}}{M_{\odot}}\right)^{\theta} M_{\oplus},
        M_{\mathrm{iso, acc}} \propto
    r^{3/20} M_{\star}^{\theta},
\end{equation}
with $\theta=-11/20$ in case of a linear scaling of the stellar accretion rate with stellar mass (Eq. \ref{eq:mdot_linear}, with $\beta$=$1$) and $\theta=23/20$, for the limit case of a quadratic scaling 
(Eq. \ref{eq:mdot_linear}, with $\beta$=$2$). 
Therefore, in the inner disc the pebble isolation mass is sensitive to the assumed accretion rate prescription \citep{liu_super-earth_2019}, but this complication does not affect lensing planets located outside the iceline (Fig. \ref{fig:pop_synth}).

\subsubsection{Planet migration}
The planetary embryos (Eq. \ref{eq:M0_pla}) grow through pebble accretion  and migrate through the disc at the same time. As they are still embedded in the disc, the asymmetric torque due to Lindblad resonances and co-rotation leads to inwards type I migration \citep{goldreich_disk-satellite_1980, ward_protoplanet_1997}. As they grow more massive, the embryos perturb the gas in the disc, creating a gap and reducing their migration rate, falling into the slower type II migration  \citep{lin_planetary_1986}, that we modelled according to \citet{kanagawa_radial_2018}. 

The inwards migration of the planet is halted at the inner edge of the disc, represented by the magnetospheric cavity 
as in \citet{frank_accretion_2002}, \citet{armitage_astrophysics_2010}, \citet{liu_dynamical_2017}
\begin{multline}
    \label{eq:r_mag_cav}
    r_{\mathrm{mag, cav}} = \left(\frac{B_{\star}^4 R_{\star}^{12}}{4GM_{\star}\Dot{M}_{\star}^2}\right)^{1/7}  \\ 
     \simeq 0.0167 \: \left(\frac{B_{\star}}{1 \: \rm kG}\right)^{4/7} \left(\frac{R_{\star}}{R_{\odot}}\right)^{12/7} \left(\frac{M_{\star}}{M_{\odot}}\right)^{-1/7} \left(\frac{\Dot{M}_{\star}}{10^{-8} M_{\odot}/ \rm yr}\right)^{-2/7} \rm AU, 
\end{multline}
where $B_{\star}$ is the stellar magnetic field, $R_{\star}$ the stellar radius, $M_{\star}$ the stellar mass and $\Dot{M}_{\star}$ the accretion rate on the central star (see Section \ref{sect:model}).
 We set our nominal values to $B_{\star} = 1 \: \rm kG$, representative of a typical solar mass T-Tauri star \citep[e.g.][]{Johns-Krull_2007}.
To model the relation between star radius and mass, we adopt the empirical relation \citet{demircan_stellar_1991}
\begin{equation}
    \label{eq:R_star}
    \log \left(\frac{R_{\star}}{R_{\odot}}\right) = 0.003 + 0.724 \log \left(\frac{M_{\star}}{M_{\odot}}\right).
\end{equation}

\section{Model calibration and variations}
\subsection{Fiducial model calibration}
\label{app:nominal_model_calibration}
\begin{figure*}[h!]
    \centering
    \includegraphics[width=\textwidth]{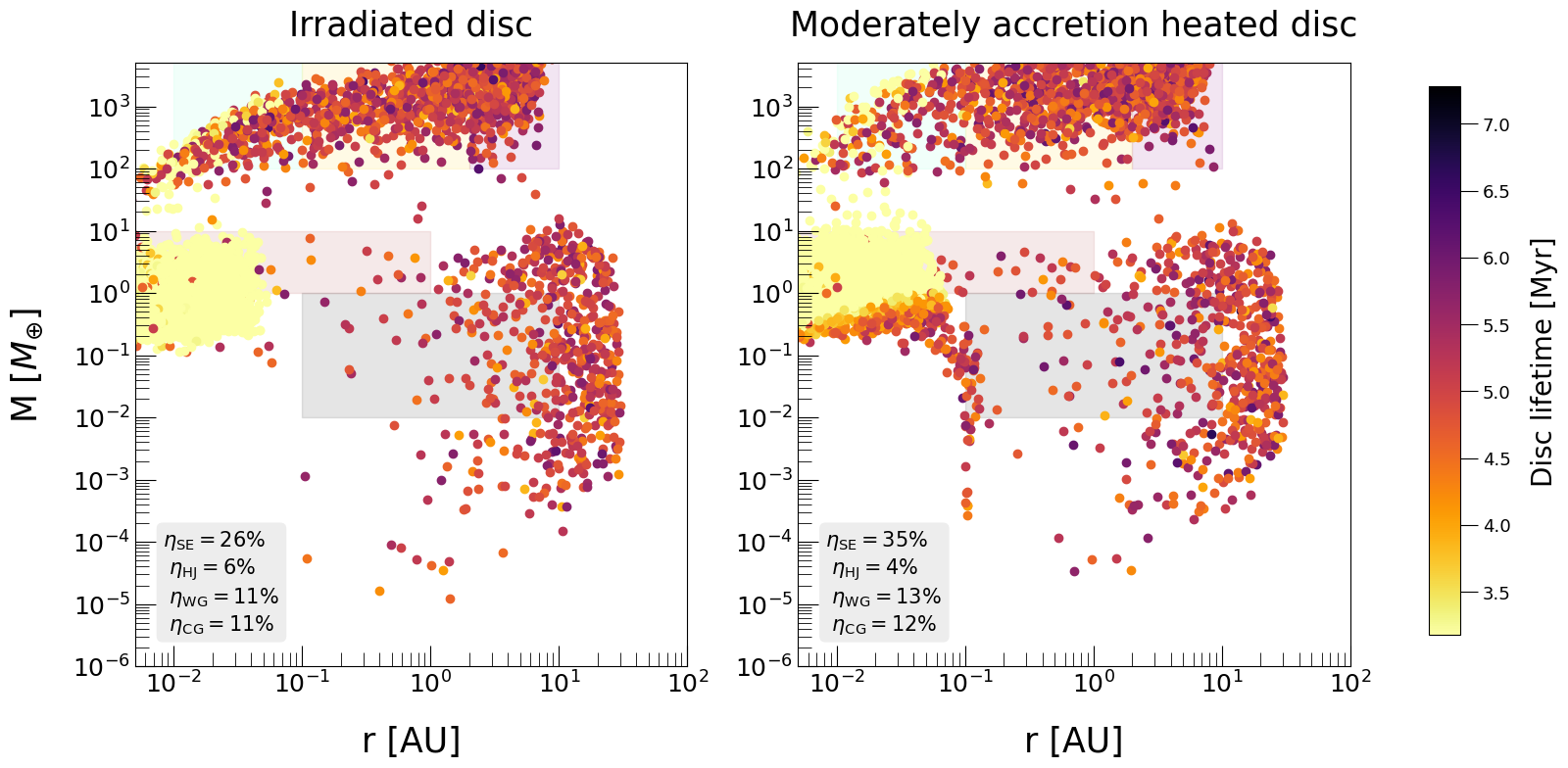}
    \caption{\textbf{Left panel}: Population synthesis of 5000 planets for an irradiated disc, with linear scaling with stellar mass (Eq. \ref{eq:mdot_linear}, $\beta = 1$), random sampled $Z$ and disc lifetime from Gaussian distributions Section \ref{app:nominal_model_calibration}) and $M_{\star}$ from the \citet{Chabrier_2005} IMF restricted between $0.01 -5 \: M_{\odot}$. The colour coding shows the disc lifetime. \textbf{Right panel}: same as left panel, but for a moderately accretion-heated disc. The bottom left box shows the occurrence fraction of the different types of planets in both cases. 
    We chose to use as a fiducial model the moderately accretion-heated disc (right panel) as the occurrence rate of super-Earths is broadly consistent with observations \citep{zhu_about_2018}.}
    \label{fig:nominal_model}
\end{figure*}
\begin{table}
    \caption{Planet type definitions}
    \label{tab:planets_def}
\centering
    \begin{tabular}{ ccc }
        \hline
        \hline
        Planet type &  Mass [M$_{\oplus}$] & Orbital distance  \\
        \hline
        Hot Jupiters (HJ) & $100 < M_{\rm p}< 6000$ & $a_{\rm p} < 0.1 \: \rm AU$ \\
        Warm Jupiters (WJ)& $100 < M_{\rm p}< 6000$ & $0.1<a_{\rm p} < 2 \: \rm AU$ \\
        Cold Giants (CG)  & $100 < M_{\rm p}< 6000$ & $a_{\rm p} > 2 \: \rm AU$ \\
        Super-Earths (SE) & $1 < M_{\rm p}< 10$     & $a_{\rm p} < 1 \: \rm AU$ \\
        Sub-Earths (sub-E)& $0.01 < M_{\rm p}< 1$   & $0.1<a_{\rm p} < 10 \: \rm AU$ \\
       \hline
    \end{tabular}
\end{table}

We calibrate our nominal model in order to roughly reproduce the occurrence rate of super-Earths (SE) around FGK stars, $P(\rm SE) \approx 30-50 \%$ \citep{zhu_about_2018}.
To find the best fit model, we simulate $5000$ planets in a purely irradiated disc (left panel of Figure \ref{fig:nominal_model}) and a moderately accretion-heated disc (right panel of Figure \ref{fig:nominal_model}), using a linear gas accretion rate scaling (see Section \ref{sect:pop_synth_stars} and Figure \ref{fig:pop_synth}). The initial positions of the planetary embryos are randomly drawn from a log-uniform distribution between $0.1$ and $30$ AU, and the initial masses are taken from the top of the streaming instability (Eq. \ref{eq:M0_pla}).
The stellar masses are randomly sampled from the \citet{Chabrier_2005} IMF distribution, limited in the range $0.01 - 5 \: M_{\odot}$. 
The stellar metallicities are sampled from a Gaussian distribution peaked at $\mu_{\rm [Fe/H]} = -0.02$ with spread $\sigma_{\rm [Fe/H]} = 0.22$ \citep{Emsenhuber_2021}.
The initial dust-to-gas ratio of the disc, $Z$, is then calculated from the star metallicity by means of $Z = 10^{\rm [Fe/H]} f_{\rm dtg, \odot}$, with $f_{\rm dtg, \odot} = 0.0149$ following \citet{Lodders_2003}. 
The accreted pebbles are both fragmentation and drift-limited in size \citep{brauer_coagulation_2008}. The value of the fragmentation velocity changes from $10$ m/s to $1$ m/s when the pebbles cross the iceline, which is defined as the location in the disc where the temperature is $T=170 \: \rm K$.
The disc viscosity is fixed at $\alpha_{\nu} = 10^{-2}$ \citep{appelgren_disc_2023}, while the turbulent stirring and fragmentation parameters are both fixed at $\alpha_{z} = \alpha_{\rm frag} = 10^{-4}$, representing a low-turbulence midplane supported by recent dust scale heights observations \citep{pinte_2016} and by observational estimates of fragmentation-limited particle sizes.
We random sample disc lifetimes, $\tau_{\rm disc}$, from a Gaussian distribution with mean $\mu_{\tau} = 5$ Myr and standard deviation $\sigma_{\tau}=0.5$ Myr.
This results in a solid disc mass distribution with median mass of $M_{\rm dust, tot} \approx 500 \: M_{\oplus}$ for a linear scaling of gas accretion rate with stellar mass (Eq. \ref{eq:mdot_linear}, $\beta=1$) and $M_{\rm dust, tot} \approx 240  \: M_{\oplus}$ for super-linear scaling (Eq. \ref{eq:mdot_manara}). 
We verified that this randomly sampled initial dust disc mass distribution is consistent with that inferred from observations of discs around young stars \citep{Tychoniec_2020} and population synthesis models of pebble disc evolution \citep{appelgren_disc_2023}.

Figure \ref{fig:nominal_model} shows the comparison between the population of planets for an irradiated disc (left) and for a moderately accretion-heated disc (right), colour-coded for disc lifetime. The occurrence rates of each type of planet in Table \ref{tab:planets_def} are listed in the bottom left corner. The moderately accretion-heated model provides a better fit to the observed super-Earth exoplanet population, therefore we chose this as the nominal disc model.

\subsection{Disc metallicity}
\begin{figure}[t!]
    \centering
    \includegraphics[width=0.86\linewidth]{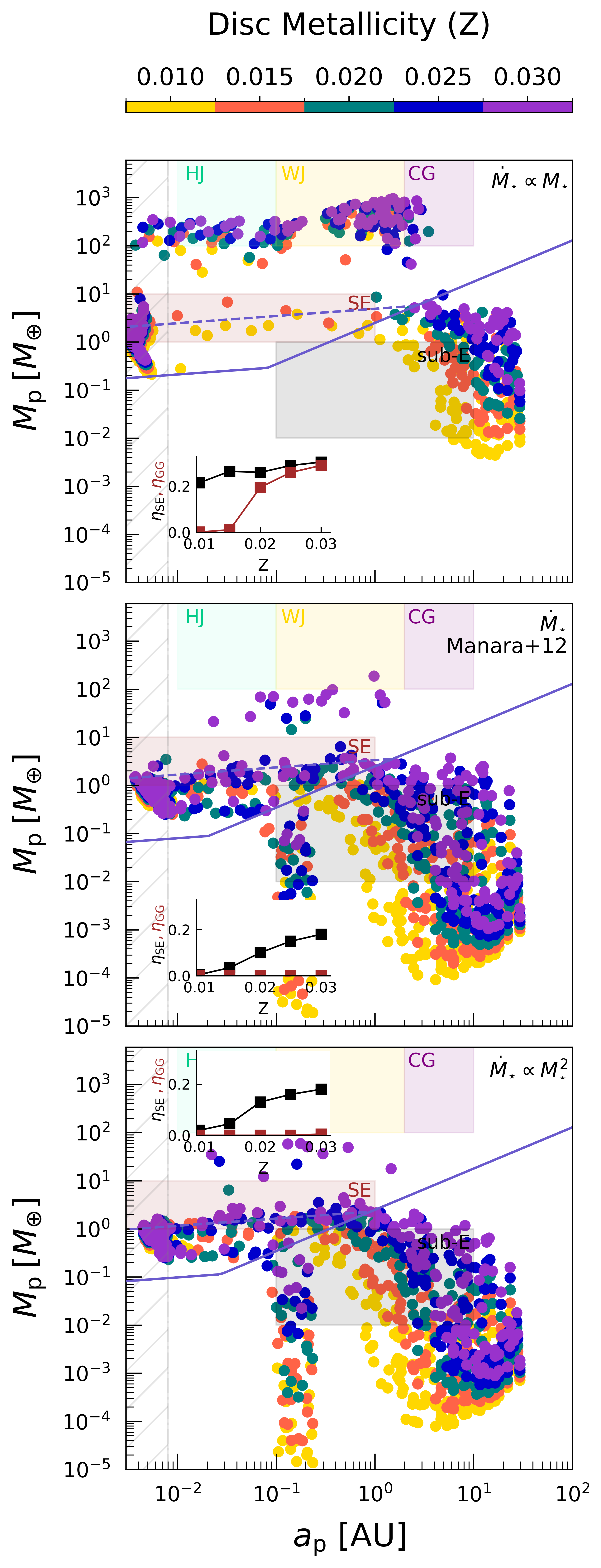}
    \caption{Semi-major axis and masses of a synthetic population of planets orbiting a $0.2 \: M_{\odot}$ star, colour coded for initial disc metallicity. The three panels correspond to the three different gas accretion rates (top panel Eq. \ref{eq:mdot_linear} with $\beta =1$, central panel Eq. \ref{eq:mdot_manara} and bottom panel Eq. \ref{eq:mdot_linear} with $\beta =2$). The purple dahsed and solid line represent the pebble isolation mass at initial and final simulation time, respectively. The inset plots show the occurrence rate of super-Earths (black lines) and giant planets (brown lines) as a function of initial disc metallicity. 
    }
    \label{fig:fixed_Z_surf}
\end{figure}
To investigate how the initial disc metallicity affects the planets final semi-major axis and mass distribution we performed a set of single planet simulations for a fixed $M_{\star} = 0.2 \: M_{\odot}$ star, using the nominal accretion-heated disc. The initial conditions of the simulations are the same as described in the model calibration of Section \ref{sect:model}, with the exception of the fixed stellar mass and the fixed initial disc metallicity from $Z =0.010, 0.015, 0.020, 0.025, 0.030$.
The three panels of Figure \ref{fig:fixed_Z_surf} show the final semi-major axes and masses of the planet populations for the three different gas accretion rates: linear dependence in the top panel (Eq. \ref{eq:mdot_linear} with $\beta =1$), steeper-than-linear for the central panel (Eq. \ref{eq:mdot_manara}) and quadratic dependence for the bottom panel (Eq. \ref{eq:mdot_linear} with $\beta = 2$). The colour-coding represents the fixed initial disc metallicity $Z$. Finally, the inset plots show the occurrence rates of super-Earths and giant planets (black and brown lines, respectively).

Simulations with higher disc metallicity show a clear increasing trend of giant planet occurrence, when assuming a linear gas accretion rate (top panel of Figure \ref{fig:fixed_Z_surf}, brown line of the inset plot), as already supported by multiple theoretical \citep[e.g.][]{Pan_2025} and observational studies \citep{Mulders_2015, van_der_marel_stellar_2021}. A steeper-than-linear gas accretion rate with stellar mass (central and bottom panels of Figure \ref{fig:fixed_Z_surf}), leads to almost no gas giant formation as the mass budget and the pebble accretion efficiency are too low regardless of the initial disc metallicity.
The super-Earths occurrence rate shows a flat trend with increasing initial disc metallicity in case of the linear gas accretion rate (top panel of Figure \ref{fig:fixed_Z_surf}, black line of the inset plot), while it increases for steeper-than-linear stellar mass dependencies (central and bottom panel of Figure \ref{fig:fixed_Z_surf}), with significant occurrence only in high metallicity discs with $Z>0.02$. This is due to the fact that the solid mass budget is lower than in the linear case, therefore a higher disc metallicity provides the needed increase in mass for more efficient super-Earth formation. 
In the case of lower-mass planets (sub-Earths), a higher intial disc metallicity leads to slightly more massive planets.

\subsection{Multiplicity: 4 planets population}
\label{app:multiple_planets}
\begin{figure}[t!]
    \centering
    \includegraphics[width=\linewidth]{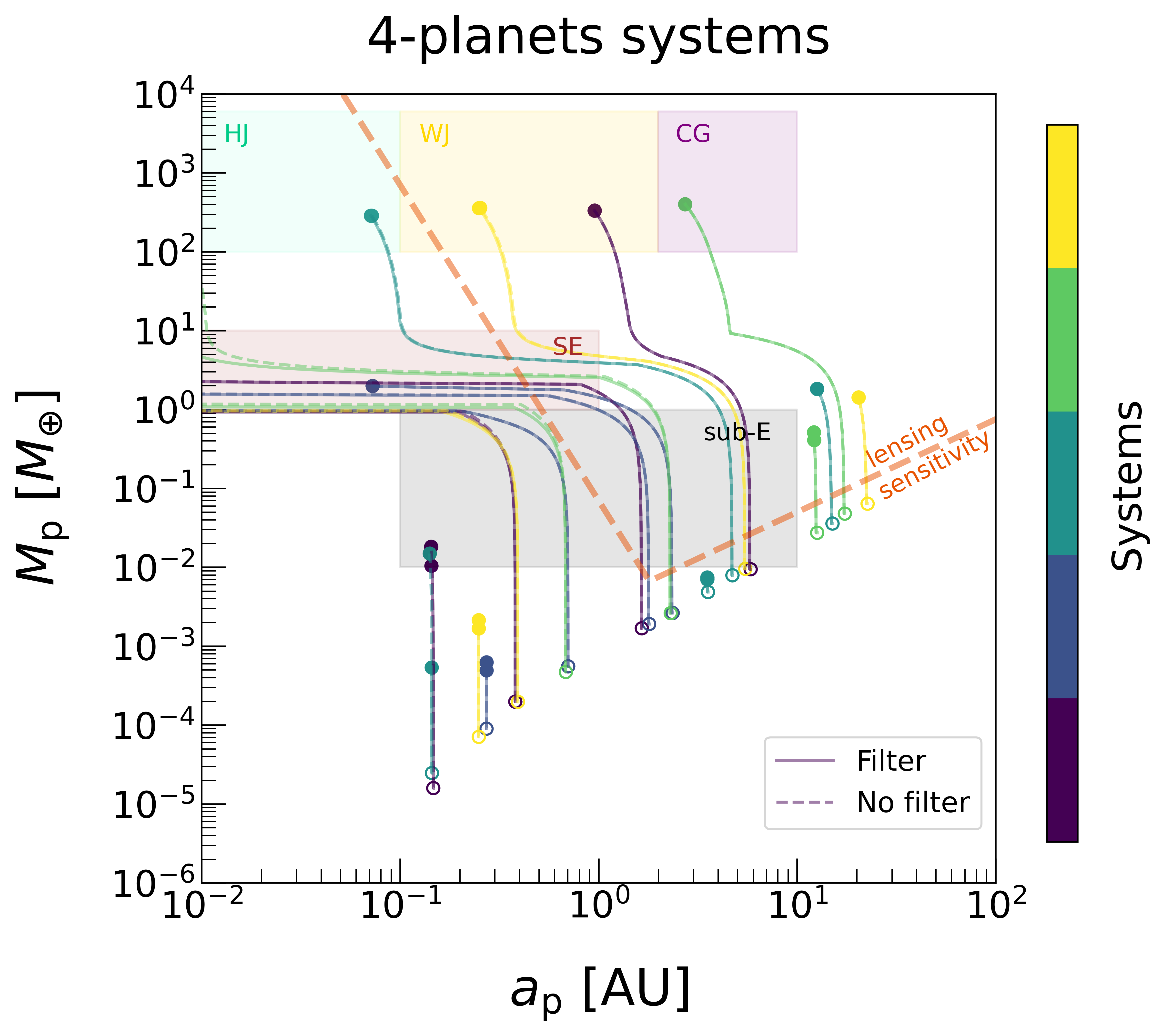}
    \caption{Growth tracks of five systems of four planets (in different colours). The solid lines show the simulations where mutual pebble filtering is accounted for, while the dashed lines show the ones without filtering.}
    \label{fig:4planets}
\end{figure}
We investigated the role of mutual filtering between embryos by running systems of $4$ planetary embryos each, randomly sampled from a log-uniform distribution in semi-major axis between $0.1$ and $30$ AU, and using a fixed disc lifetime of $5$ Myrs.
Figure \ref{fig:4planets} shows example growth tracks of the planets in our 4-planet system simulations, marked by a solid line when pebble filtering is included and by a dashed line when pebble filtering is not included. Each colour identifies a different randomly-drawn system of 4 planets. The lensing sensitivity triangle is marked by the orange dashed lines.

We find that the occurrence rate of lensing planets is hardly affected by the multiplicity of the system. Mutual filtering is irrelevant in the lensing corner, at most leading to slightly lower masses in the inner disc (see the teal innermost planet in Figure \ref{fig:4planets}). 

\section{The lensing sensitivity triangle}
\label{app:triangle_efficiency}
\begin{figure*}[t!]
    \centering
    \includegraphics[width=\textwidth]{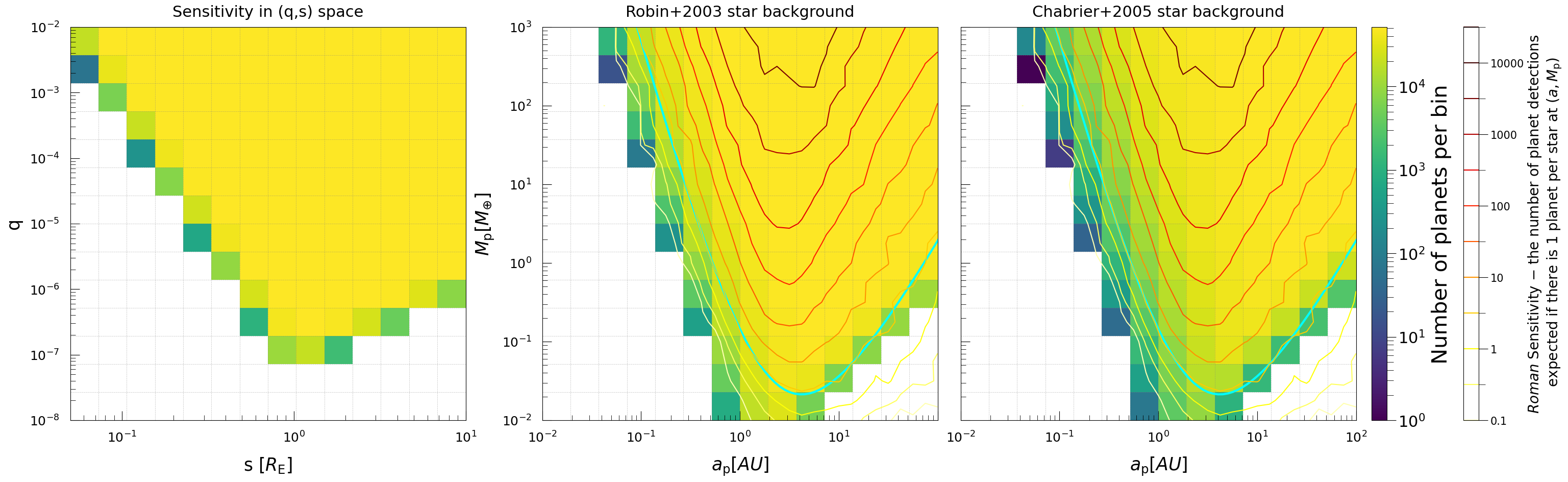}
    \caption{\textbf{Left panel}: Sensitivity in the $(q,s)$ space for a log-uniform distribution of synthetic planets.
    \textbf{Central and right panel}: Detected planets from a log-uniform distribution of synthetic planets in ($M_{\mathrm{p}}, a_{\mathrm{p}}$), for two different star initial mass functions: left panel for the \citet{Robin_2003} disc stars power law and right panel for the \citet{Chabrier_2005} IMF. The colour-coding is the total number of detected stars in each bin. The contours show the detection sensitivity lines of \textit{Roman} from \citet{Penny_2019}, with the $3$-detection line in cyan, assuming one planet per star.}
    \label{fig:sensitivity_star_IMF}
\end{figure*}
The microlensing sensitivity calculated through Equation\,\ref{eq:s_q} results in a triangle-shaped detection efficiency in the $(q,s)$ space, which has been observed to nicely reproduce most of the high magnification events \citep{gould_2010}. 
The tip of the sensitivity triangle, $q_{\mathrm{min}}$, which depends on the specifics of the instrument and on the microlensing event, is parametrised as
\begin{equation}
    \label{eq:qmin}
    q_{\rm min} = \frac{\xi}{A_{\rm max}},
\end{equation}
where $\xi \approx 1/50$ is a parameter that depends on the quality of data, while $A_{\rm max}$ is the maximum magnification. Here, $1/A_{\rm max}$ essentially probes how close the source comes to the central caustic of the system \citep{gould_2010}

Parametrising the survey sensitivity with a triangle of $100 \%$ detection efficiency inside and $0 \%$ outside is a reasonable approximation for high-magnification events, which are nearly symmetrical in $\log s$ \citep{gould_2010, choi_2012b}, but overestimates the detection efficiency in case of mid and low-magnification events, that show higher sensitivities outside the Einstein ring ($s>1$) because the magnification is the result of the planetary perturbation of the major image which is brighter \citet{suzuki_microlensing_2016}. In case of a primary event with magnification $A$, the perturbation of the major image results in a magnification $(A+1)/2$, while the perturbation of the minor image is $(A-1)/2$, therefore, as $A \rightarrow1$, the only observable magnification events are the ones that perturb the major image ($s>1$).
The sensitivity for low magnification events also tends to be lower at exactly the Einstein separation, leading to a W-shaped rather than triangular-shaped detection efficiency \citep[e.g.][]{suzuki_microlensing_2016}.
In principle, the planet detection efficiency is not only a function of ($q,s$) but also $\alpha$, the angle of the source–lens trajectory relative to the planet–star axis, but the narrowness of the boundary region between $100 \%$ and $0 \%$ detection efficiency suggests it to be nearly independent of $\alpha$ \citep{Batista_2009,gould_2010}, hence we do not parametrise it.

In order to estimate the number of detected planets from our underlying planet distribution, we mapped $(M_{\mathrm{p}}, a_{\mathrm{p}})$ into the $(q,s)$ space to check the detectability of the planet, making use of
\begin{align}
    \label{eq:M_q}
    M_{\mathrm{p}} & = q M_{\mathrm{L}}, \\
    \notag
    a_{\mathrm{p}} & = s R_{\mathrm{E}},
\end{align}
where $M_{\mathrm{L}}$ is the lens (host) star mass and $R_{\mathrm{E}}$ the Einstein radius
\begin{align}
    \label{eq:R_E}
        R_{\rm E} & = \theta_{\rm E} D_{\rm L} 
        \notag\\
        &
        = 2.85 \: \rm AU \left(\frac{M_{\rm L}}{0.5 M_{\odot}}\right)^{1/2} \left(\frac{D_{\rm S}}{8 \rm kpc}\right)^{1/2} \left(\frac{D_{\rm L}/D_{\rm S}(1-D_{\rm L}/D_{\rm S})}{0.25}\right)^{1/2}\,.
\end{align}
Here $D_{\rm L}$ and $D_{\rm S}$ are the lens and source star distances, whose nominal values, assuming they are a disc and bulge star respectively, are $4$ and $8$ kpc.  
The Einstein ring is defined as
\begin{equation}
    \label{eq:theta_E}
    \theta_{\rm E} = \left(\frac{4GM_{\rm L}}{c^2 D_{\rm LS}}\right)^{1/2},
\end{equation}
where $D_{\rm LS}$ is the lens to source distance and $c$ is the speed of light.

As we see from the above expressions, the $(q,s) \rightarrow (M_{\mathrm{p}}, a_{\mathrm{p}})$ mapping requires knowledge of the lens (host) star mass. 
When dealing with our synthetic population of planets, we know the star's masses as we draw them from a given initial mass function distribution (for Figure \ref{fig:lensing}, based on the \citet{Chabrier_2005} IMF).
Also, for real observed lensing exoplanets, the uncertainty on the stellar mass introduces an important uncertainty on the planet's properties.
Estimates of the host star mass are possible if the event is long enough to measure the effect of the annual microlensing parallax or if it is observed from two widely separated observers, which proves difficult for microlensing surveys, especially from the ground. High-resolution imaging provides an alternative method to estimate the host star's mass, by measuring the separation between source and host star and the host star's colour and magnitude. The \textit{Roman} Space Telescope will be able to make these measurements routinely for most of the events \citep{Penny_2019}. It is indeed expected that \textit{Roman} will be able to determine the lens star masses and distances with uncertainties of less than $20 \%$ for half of the events and over $40 \%$ for all of them \citep{Terry_2025}. 

The left panel of Figure \ref{fig:sensitivity_star_IMF} shows how we parametrised the lensing sensitivity in the $(q,s)$ space throughout this paper, where we approximated a detectability of $100\,\%$ inside the sensitivity triangle and $0 \%$ outside. The parameters that define the triangle through Eq. \ref{eq:s_q} have values $q_{\rm min} = 10^{-7}$, $\eta_+=0.85$ and $\eta_-=0.25$ to roughly match the \textit{Roman} sensitivity.
The central and right panels of Figure \ref{fig:sensitivity_star_IMF} show an example of the detectability of a sample of synthetic planets drawn from a log-uniform distribution in mass and semi-major axis (same as the central panel of Figure \ref{fig:lensing}), assuming two different underlying stellar IMFs. The central panel draws from the \citet{Robin_2003} stellar IMF, while the right panel draws from the \citet{Chabrier_2005} IMF.
The colour-coding represents the number of detected planet per each bin, which, in the underlying log-uniform synthetic planet distribution, are $\approx$\,$50\times 10^3$ planets per bin.
For comparison, the coloured contours depict the \textit{Roman} sensitivity detection lines from \citet{Penny_2019}, assuming one planet per star, with the $3$-detection line marked in light blue.
Given our $100 \%$ detectability inside and $0 \%$ outside the triangle, we tend to generally overestimate the detectability of planets, as shown from the comparison to the \textit{Roman} sensitivity contours in the central and right panel of Figure \ref{fig:sensitivity_star_IMF}.
Nonetheless, the break of planet occurrence near the iceline (right panel of Figure \ref{fig:lensing}) would hold even with a more accurate sensitivity estimation, as it appears in a region where the sensitivity of microlensing from space is reasonably close to unity.

The comparison between the central and right panel of Figure \ref{fig:sensitivity_star_IMF} also shows how the estimate of the mass of the lensing star
affects our ability to detect a planet with a certain mass $M_{\mathrm{p}}$ orbiting at a certain distance $a_{\mathrm{p}}$. As the planetary caustics depend on the planet-to-star mass ratio, the same type of planet orbiting a different type of star might not be visible. The \citet{Chabrier_2005} stellar IMF from the right panel is peaked around $0.2 \: M_{\odot}$, while the central panel IMF \citep{Robin_2003}, peaks around less massive stars. The lensing sensitivity in the right panel decreases more gradually for lower $q$ and further away from the Einstein radius. 
This happens because for a fixed planet mass, $M_{\mathrm{p}}$, orbiting at a fixed distance, $a_{\mathrm{p}}$, an increase in the mass of the host star $M_{\mathrm{L}}$ implies a lower mass ratio $q$ and a higher projected separation $s$ (Eq. \ref{eq:s_q}), therefore effectively pushing the planet outside of the detection sensitivity triangle.

\end{appendix}

\end{document}